# An experimental and numerical study of water jet cleaning process

Anirban Guha[a,1,*], Ronald M. Barron[a,b], Ram Balachandar[c]

[a] Mechanical, Automotive & Materials Engineering Department, University of Windsor
 Windsor, Ontario Canada N9B 3P4
[b] Department of Mathematics & Statistics, University of Windsor
 Windsor, Ontario Canada N9B 3P4
[c] Civil and Environmental Engineering Department, University of Windsor
Windsor, Ontario Canada N9B 3P4

[*] Corresponding author Tel: +1 604 345 8951; email: aguha@interchange.ubc.ca
[1] Present Address: Civil Engineering Department, The University of British Columbia, Vancouver, BC CanadaV6T 1Z4

## ABSTRACT

In this paper, we have experimentally, numerically and theoretically investigated the water jet cleaning process. Very high speed water jets (~80-200 m/s) are typically used in such cleaning operations. These jets diffuse in the surrounding atmosphere by the process of air entrainment and this contributes to the spreading of the jet and subsequent decay of pressure. Estimation of this pressure decay and subsequent placement of the cleaning object is of paramount importance in manufacturing and material processing industries. Also, the pressure distribution on the cleaning surface needs to be assessed in order to understand and optimize the material removal process. In an industrial setting, experimental study is performed to formulate the pressure characteristics. It has shown that the decay of stagnation pressure along the axial direction is linear. It also revealed that no cleaning is possible at radial locations greater than 1.68$D$ from the centerline. Numerical simulations are performed to capture the process of air entrainment in the jet and the subsequent pressure characteristics. The simulation results are found to correctly predict the experimental data. Moreover, a theoretical model for evaluating the optimal and critical stand-off distances has also been derived. Based on our results, we found that the optimal stand-off distance in cleaning operations is ~5$D$ and the jet looses its cleaning ability at axial distances greater than ~26$D$.

# 1. Introduction

High speed water jets in air are extensively used in manufacturing industry for cutting and cleaning operations. Water jets are used for removal of various coatings or deposits from the substrates and also for the cutting of many materials. While water jet cutting involves the penetration of a solid material by a continuous jet, water jet cleaning involves an erosion process by which deposits are removed from the material surface. Hashish and duPlessis (1978,1979) investigated the jet cutting process both analytically and experimentally. They performed a control volume analysis to evaluate the hydrodynamic forces. Later, their work was extended to find the optimal stand-off distance. Similar analysis was performed by Leu et al. (1998) for the case of water jet cleaning. Cleaning water jets generally have the velocity range of ~80-200 m/s. They exhibit a high velocity coherent core surrounded by an annular cloud of water droplets moving in an entrained air stream. Leu et al. (1998) as well as Rajaratnam et al. (1994, 1998) have discussed the anatomy of high speed water jets in air; see Figure 1. Such jets can be divided into three distinct regions:

a) *Potential Core Region*: This region is the one close to the nozzle exit. In this region, primary as well as secondary Kelvin-Helmholtz instabilities bring about transfer of mass and momentum between air and water. The process of air entrainment breaks up continuous water into droplets. There remains a wedge shaped potential core surrounded by a mixing layer in which the velocity is equal to the nozzle exit velocity.

b) *Main Region*: The continuous interaction of water with surrounding air results in the break up of the water jet stream into droplets; the size of which decreases with the increase of radial distance from the axis. Since the jet transfers momentum to the surrounding air, its mean velocity decreases and therefore it expands. The region closest to the jet-axis is known as the water droplet zone. There is another zone; viz. the water mist zone, which separates the droplet zone from the surrounding air. This mist zone is characterized by drops of very small size and negligible velocity.

c) *Diffused Droplet Region*: This zone is produced by the complete disintegration of the jet into very small droplets having negligible velocity.

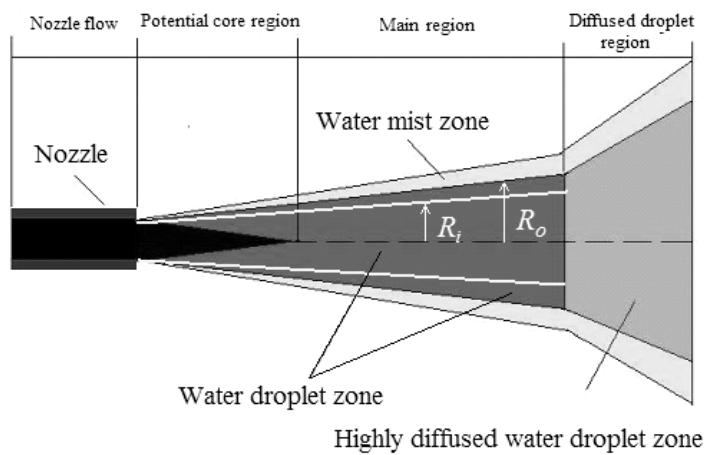

Figure 1: Anatomy of high speed water jets in air

Rajaratnam et al. (1994, 1998) did experiments with a converging-straight nozzle having nozzle exit diameter, $D$=2 mm and nozzle exit velocity ($V_o$) around 155 m/s. They found that the centreline velocity of the jet remains constant and equal to $V_o$ for more than 100$D$ and then decays linearly to about 0.25 $V_o$ at about 2500$D$. Surprisingly; severe air entrainment causes the water volume fraction ($α_w$) to fall drastically. Measurements along the centreline show that $α_w$ at 20$D$ is 20%, at 100$D$ is 5% and at 200$D$, it is just 2%. Although they didn't perform any pressure measurements, it is evident that such high degree of air entrainment will drastically decay the centreline pressure. The entrained air flows with the same velocity as the water phase, but owing to the very low density of air in comparison to water (~1:815); the momentum delivered to the cutting or cleaning surface will be significantly reduced. Thus at moderately large stand-off distances (distance of the target surface from the nozzle exit measured along the centreline), efficient cleaning will be impossible.

Research on numerical simulations of high speed water jets has been very limited; and perhaps unsuccessful. Liu et al. (2004) tried to simulate the abrasive water jet (AWJ) cutting problem, but their results were not validated against experimental results. They used VOF multiphase model in FLUENT to simulate the two phase (water-air) flow but unfortunately, this model (or any multiphase model) is *not* able to capture the air entrainment process. This is why the results of Liu et al. fail to simulate the actual flow physics as experimentally observed by Rajaratnam et al. (1994, 1998). It is worth mentioning here that numerical simulation of high speed jets faces significant challenges. Lin et al. (1998) concluded that the process of jet break-up and subsequent mist formation not only depends on the thermodynamic states of both the liquid jet and the ambient air but also has a strong dependence on the nozzle internal flow characteristics, e.g. nozzle cavitation, turbulence, etc. Yoon et al. (2004) did a series of experiments with low, moderate and high speed jets. They inferred that the process of hydrodynamic instability and subsequent break-up mechanism of high speed jets (Weber Number greater than 100,000) is significantly different from the better understood Rayleigh type mechanisms of the low/moderate speed jets. This infers that the theoretical understanding of mechanisms related to high speed water jets is still not well developed.

To bypass the theoretical limitations, Guha et al. (2010) developed a semi-empirical model to capture the process of air entrainment. Based on empirical relations, this model evaluates the interaction between air and water phases. This interaction term is then incorporated into the governing Navier-Stokes equations as source term. They validated their simulations against the experimental results of Rajaratnam et al. (1994, 1998).

The primary objective of this paper is to contribute to the knowledge of high speed jet impingement on a target plate and its effect on material removal process. Before discussing any further, a brief introduction to the material removal process is necessary. Jet impingement on the target coating creates impact forces. The mechanism of material removal is very complex. Adler (1979) proposed that material erosion by the water jets consist of four damage modes, viz. direct deformation, stress wave propagation, lateral outflow jetting, and hydraulic penetration. One or more damage modes may exist in a particular erosion process. Direct deformation and stress wave propagation are responsible for initiation of cracks. Watson et al.(1993) and Springer (1976) concluded that the propagation of stress waves caused by the impact forces is generally responsible for crack initiation in the erosion process. Lateral outflow jetting and hydraulic penetration cause the extension, enlargement and propagation of existing cracks. In the erosion of coating material, the adhesion between the coating and the substrate may also need to be considered.

In order to have thorough understanding on the erosion process and thereby to produce an efficient cleaning, it is essential to have knowledge on the magnitude and distribution of the water jet pressure on the target plate. Leach et al. (1966) studied the pressure distribution on a target plate placed at a given axial distance from the nozzle. They found that although the normalized pressure distribution along the centreline of a jet is dependent on the nozzle geometry, it is independent along the radial direction. Their investigations showed that the jet pressure becomes equal to the ambient pressure at a distance of around 1.3$D$ from the centreline. Outside this region, the shear stress is too small to perform cleaning of the target surface. They also found that the normalized pressure distribution was similar for different inlet pressure conditions as well as different nozzle geometries.

In this paper, we have performed experimental as well as numerical studies to investigate the high speed water jet cleaning process. In an industrial setting, we have performed experiments to capture the centreline pressure as well as the radial distribution of pressure on the cleaning surface. For numerical simulations, we have used the semi-empirical model of Guha et al. (2010) to capture the air entrainment process. It is worthwhile to mention that previously the model of Guha et al. has been mainly validated against the results of a free jet. Here we have extended the model for the case of an impinging jet. Thus, the model has been thoroughly examined against our experimental results capturing the pressure characteristics on the target plate. Also we have used our experimental data to develop an empirical formula on the magnitude and distribution of pressure on the target plate at a given stand-off distance. Apart from that, we have made a brief contribution to the theoretical aspect of water jet impingement by deriving formulas for optimal and critical stand-off distances.

## 2. Semi-Empirical mass-flux model

The semi-empirical mass flux model of Guha et al. (2010) combines analytical and empirical methods to evaluate the mass flux at a particular location ($x$, $r$) inside the jet. It is expressed as follows:

$$\dot{M}(x,r) = \frac{5.62 \times \rho_w \times \alpha_{wo} \times V_{wo} \times R_N^2}{R^2} \left\{ 1 - \left(\frac{r}{R}\right)^{1.5} \right\}^3 \quad (1)$$

Here, $\dot{M}(x,r)$ is the mass flux of water droplets at the location ($x$, $r$), $\rho_w$ is the density of water, $\alpha_{wo}$ and $V_{wo}$ are the volume fraction and axial velocity of water droplets at the nozzle exit respectively, $R_N$ is the nozzle radius. The variable $R$ is defined as

$$R = R_i \text{ if } r \leq R_i, \quad R = R_o \text{ if } R_i < r \leq R_o \quad (2)$$

Where $R_i$ is the radial width of the continuous flow region and is given by

$$R_i = k_1 \sqrt{xR_N} + R_N \qquad (3)$$

Outside of this region is the droplet flow region, the radial width of which varies as

$$R_o = Cx + R_N \qquad (4)$$

$k_1$ and $C$ are Spread Coefficients. The value of $C$ is usually in between 0.02 and 0.06. $k_1$ is estimated as follows:

$$k_1 \sqrt{R_N} = const \times C \qquad (5)$$

where $const$ is usually $O(10^{-1})$.

Schematic description of $R_i$ and $R_o$ has been provided in Figure 1.

In Equation (1), $\alpha_{wo}$ is usually 1.0. Hence, if the nozzle exit velocity and the Spread Coefficients are known, the jet flow characteristics can be properly estimated.

## 3. Numerical Simulation

Here we adopt the numerical methodology undertaken by Guha et al. (2010). In order to implement Equation (1) numerically, it is coupled with the continuity and momentum equations of turbulent multiphase flows.

### 3.1. Governing Equations and Solution Methodology

The governing equations of this problem are Navier-Stokes equations for turbulent multiphase flows. The multiphase model chosen for this problem is the Eulerian multiphase model. k-ε turbulence model with standard wall functions are used to capture turbulence.
The continuity and momentum equations for the $w$ (water) phase in the Eulerian model for multiphase flows are, respectively

$$\frac{\partial(\alpha_w \rho_w)}{\partial t} + \nabla \cdot (\alpha_w \rho_w \vec{v}_w) = \sum_{i=w,a}(\dot{m}_{a \to w} - \dot{m}_{w \to a}) + S_w \qquad (6)$$

$$\frac{\partial(\alpha_w \rho_w \vec{v}_w)}{\partial t} + \nabla \cdot (\alpha_w \rho_w \vec{v}_w \vec{v}_w) = -\alpha_w \nabla p + \nabla \cdot \bar{\bar{\tau}}_w + \alpha_w \rho_w \vec{g} +$$

$$\sum_{i=w,a}\{K_{wa}(\vec{v}_w-\vec{v}_a)+\dot{m}_{a\to w}\vec{v}_{a\to w}-\dot{m}_{w\to a}\vec{v}_{w\to a}\}+\vec{F}_w \tag{7}$$

The term $\dot{m}_{w\to a}$ is the mass transfer from $w$ (water) phase to $a$ (air) phase. In order to simulate the air entrainment process, we set $\dot{m}_{a\to w}$ and $S_w$ as zero and

$$\dot{m}_{w\to a}=\nabla\cdot(\dot{M},0) \tag{8}$$

where $\dot{M}$ is known from Equation (1).

FLUENT is used as the flow solver. Water is treated as the secondary phase. The drag coefficient between the two phases is determined by the Schiller-Naumann equation (Fluent 6.3.26 User Manual). Equation (8) is incorporated into FLUENT by the means of User Defined Functions (UDF). The source term due to momentum transfer ($\dot{m}_{w\to a}\vec{v}_{w\to a}$) in Equation (7) is automatically handled by FLUENT once the mass transfer is specified (through UDF). This is done as follows:

$$\vec{v}_{w\to a}=\vec{v}_a \quad \text{if } \dot{m}_{w\to a}>0,\ \vec{v}_{w\to a}=\vec{v}_w \quad \text{if } \dot{m}_{w\to a}<0 \tag{9}$$

The term $K_{wa}(\vec{v}_w-\vec{v}_a)$ in Equation (7) represents inter-phase interaction force and $K_{wa}$ is the inter-phase momentum exchange coefficient.

The *k-ε* mixture turbulence model is used for turbulence modeling. The transport equations are as follows:

$$\frac{\partial(\rho_m k)}{\partial t}+\nabla\cdot(\rho_m k \vec{v}_m)=\nabla\cdot\left(\frac{\mu_{t,m}}{\sigma_k}\nabla k\right)+G_{k,m}-\rho_m\varepsilon \tag{10}$$

$$\frac{\partial(\rho_m\varepsilon)}{\partial t}+\nabla\cdot(\rho_m\varepsilon\vec{v}_m)=\nabla\cdot\left(\frac{\mu_{t,m}}{\sigma_\varepsilon}\nabla\varepsilon\right)+\frac{\varepsilon}{k}(C_{1\varepsilon}G_{k,m}-C_{2\varepsilon}\rho_m\varepsilon) \tag{11}$$

where $\rho_m$ is the mixture density and $\vec{v}_m$ is the mixture velocity. The turbulent viscosity ($\mu_{t,m}$) and the production of turbulent kinetic energy ($G_{k,m}$) are calculated as follows:

$$\mu_{t,m}=\rho_m C_\mu \frac{k^2}{\varepsilon} \tag{12}$$

$$G_{k,m}=\mu_{t,m}\left(\nabla\vec{v}_m+(\nabla\vec{v}_m)^T\right) \tag{13}$$

The values of the model constants are taken as the "standard" values $C_{1\varepsilon} = 1.44$, $C_{2\varepsilon} = 1.92$, $C_\mu = 0.09$, $\sigma_k = 1.0$, $\sigma_\varepsilon = 1.3$. Standard wall functions are used to model near wall flows. For brevity, the description of standard wall functions is not discussed. Interested readers can refer to FLUENT 6.3.26 User Manual for details.

### 3.2. Computational Domain and Boundary Conditions

Following Guha et al. (2010), the computational domain (see Figure 2) and structured grid system is created in GAMBIT. In all our problems, we have dealt with circular jets. Hence the domain is made two dimensional and axisymmetric (half of the domain was simulated). The boundary conditions appear as shown in Figure 2. The radial extent of the domain is large enough to ensure that the pressure outlet boundary condition (set at atmospheric pressure) and the wall boundary conditions can be accurately applied, i.e., without adversely affecting the flow field. It is important to note that we have used a wall boundary condition near the velocity inlet. Replacing it with the more realistic pressure outlet boundary condition may cause backflow of the jet. Also, the use of internal boundary condition helps in dividing the domain into two parts, the one inside the jet region with densely clustered grids which the other one with variable grid spacing, the size of which increases towards the pressure outlet boundary. The velocity inlet condition can be obtained by solving the nozzle flow problem separately provided the nozzle geometry and nozzle inlet conditions are known. The nozzle exit velocity profile will then become the velocity inlet condition for the jet flow problem. In all the problems considered in this paper, the nozzle exit velocity is fully developed and turbulent, and the nozzle is assumed to be frictionless. Hence the nozzle exit velocity has top-hat profile.

Firstly, we have tried to simulate the flow conditions given in Rajaratnam et al. (1994, 1998). We have made the domain size 1000 mm × 500 mm to ensure that the boundaries are far apart from the jet. Rajaratnam et al. studied free jet problem, hence the wall boundary at the opposite end of the velocity inlet is not important for this particular problem. The radial width ($R_N$) of the velocity inlet boundary (set at 155 m/s) is 1 mm as per the experimental conditions.

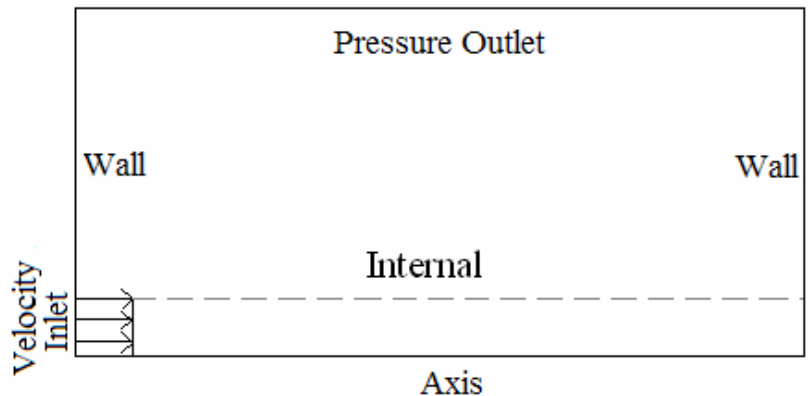

Figure 2: Computational domain and boundary conditions (Not drawn to scale)

In validating other experimental results, we use the same computational domain as shown in Figure 2, but the geometry is different and problem specific. The same is true for boundary conditions. These will be discussed in details in the relevant sections of the paper.

The initialization of the problems can be done by setting the water phase volume fraction equal to 1 at the velocity inlet and 0 elsewhere. The problem with this approach is that it will take very long time to perform the computations. To bypass this problem, we take the help of internal boundary condition shown in Figure 2. We make the water phase volume fraction inside the entire region within the internal boundary and the axis to be 1. The same is set to 0 in the other half of the domain. Since the jet will eventually reach the wall and we are interested in the steady state behavior of the problem, this initialization trick helps to save considerable amount of computational time.

In all the problems, pressure-velocity coupling is achieved using the phase-coupled SIMPLE algorithm. All the residuals tolerances are set to $10^{-6}$ and the time step size is $10^{-5}$ seconds. The program is run for a time long enough to attain quasi-steady state.

## 4. Validation of Numerical Results

Like Guha et al. (2010), we have validated the published experimental results of of Rajaratnam et al. (1994, 1998) and Leach et al. (1966). Figures 3-4 compare the simulation results with that of Rajaratnam et al.. Indeed Rajaratnam et al. found that the centreline velocity ($V_o$) of the jet remains constant for more than 100$D$ and then decays linearly to about 0.25$V_o$ at about 2500$D$. Severe air entrainment causes the water volume fraction ($\alpha_w$) to fall drastically from 20% at 20$D$ to 5% at 100$D$. Figures 3-4 confirm that the simulation accurately predicts the centreline characteristics. The User Defined Function added to the FLUENT solver for the validation of these results has been provided in the Appendix.

It is important to mention here that the radial velocity distribution (not included in study, but for reference see Guha et al. (2010)) deviates from the experimental findings of Rajaratnam et al. (1994) for radial width greater than 5$D$. At such distances, the water mist zone is more prominent. Since the mist zone is formed of sparse droplets, the continuum hypothesis becomes invalid; hence, the model is no longer suitable to capture the flow physics. Note that the mist zone has almost no effect in cleaning applications; hence its modeling is not a major concern.

From cleaning point of view, the pressure distribution on a target plate ($P_T$) placed perpendicularly to the jet flow field is of primary concern. Since the jet looses sufficient amount of centreline pressure ($P_T(x,0)$) as it progresses through the surroundings, to ensure efficient cutting or cleaning, the target plate should be kept near the nozzle exit. It is essential for the simulation to predict the pressure distribution at the target plate fairly accurately. From the previous paragraph, we came to know that the model of Guha et al. gives a good prediction of the characteristics of a free jet. Hence it has been extended to understand the pressure characteristics of an impinging jet. Thus, the experimental conditions (jet velocity of 350 m/s , nozzle radius of 0.5 mm and stand-off distance 76 mm) of Leach et al. (1966) has been numerically implemented. Here the domain is made 76mm × 250 mm to properly satisfy the geometry and boundary conditions. Care should be taken while performing the simulations with impinging jets. Near the zone of impingement, the jet dynamics changes considerably and the mass flux model no longer holds. This is because the mass flux model is based on free jets. To tackle this problem, the axial length of the jet along which the mass flux relation holds is prescribed to be a little less than that of the axial length of the domain (to be on the safe side, we kept it to be a few millimeters less than the point where stagnation took place). Figure 5 shows the comparison between the simulation results and the experiment. It seems that the numerical simulation under predicts the experimental data. The reason behind this deviation will be discussed in Section 5.

From the validations discussed above, it is evident that the mass flux model gives reasonable predictions of both the axial (centreline) as well as the radial characteristics.

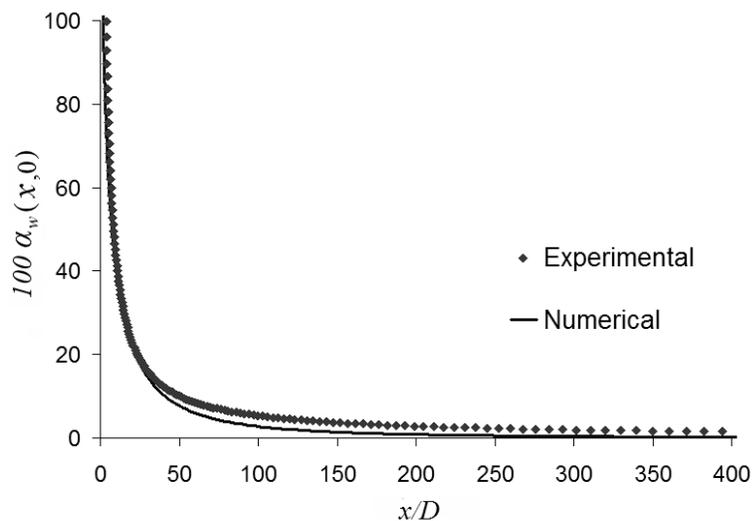

Figure 3: Numerical simulation of the decay of centreline water phase volume fraction and comparison with experimental results of Rajaratnam and Albers (1998).

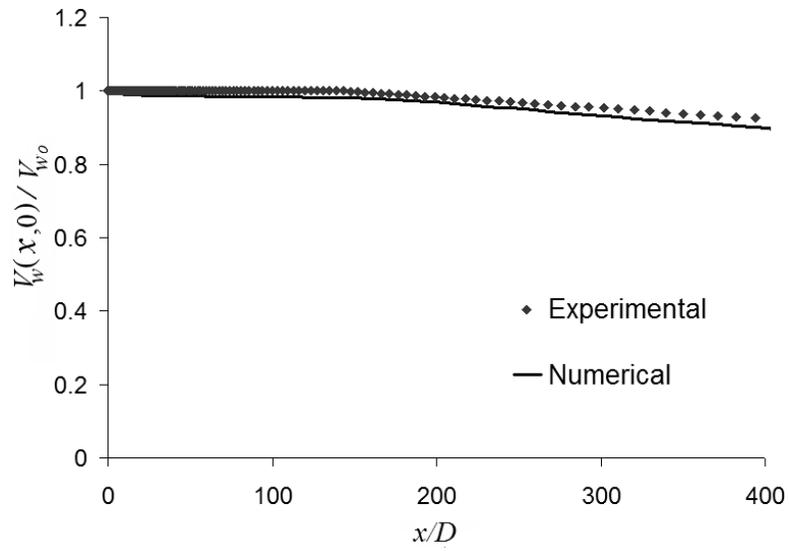

Figure 4: Numerical simulation of normalized centreline water phase velocity and comparison with experimental results of Rajaratnam et al. (1994).

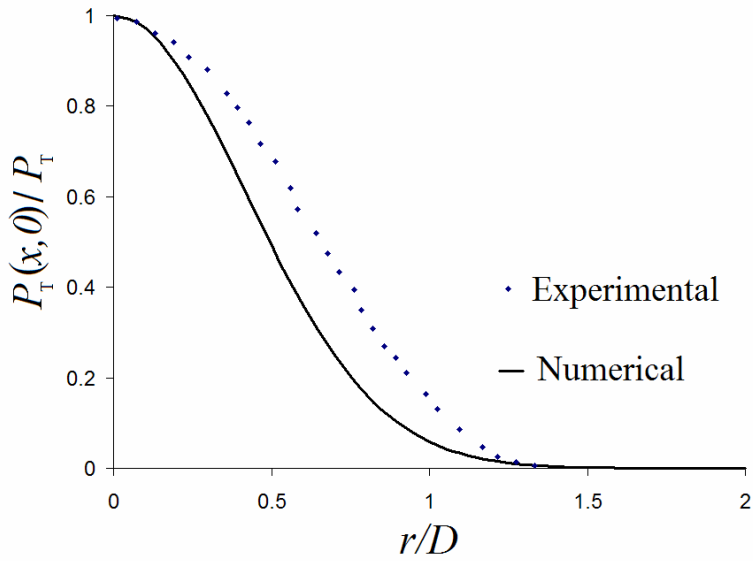

Figure 5: Normalized pressure distribution on a target plate placed at 76$D$ and comparison with Leach et al. (1966).

# 5. Experimental Set-up

As has been already discussed, there has been some experimental works on cleaning jets. Yet, there is no consummate view on the magnitude and distribution of water pressure on the target plate. Without this knowledge, it is not possible to design an efficient cleaning system. This has motivated us to perform experimental investigations on the pressure distribution on the target plate.

A schematic of the experimental setup is illustrated in Figure 6. The capacity of the pump was 0.00631 m$^3$/s and 5 MPa. The converging nozzle used in this study is 0.0457 m long, with largest diameter 0.014 m and smallest diameter ($D$) 0.0072 m. The pressure reducing valve mounted on the line feeding water from the pump to the nozzle is able to reduce the pressure to 1.03 MPa. The static pressure at the nozzle inlet is measured with the aid of a pressure transducer. The mass flow rate of water through the nozzle is measured using a collecting vessel and stop-watch. Since the nozzle geometry and mass flow rate are known, the average velocity at the nozzle exit can be easily determined. The water, flowing out of the nozzle in the form of a high speed jet, impacted onto a target, which can be moved both axially and radially with the aid of a robotic arm. The target plate has a pressure transducer (Uncertainty =2%) mounted at its centre. The distance moved by the target in both axial and radial directions is measured by a Linear Variable Displacement Transducer (Uncertainty =0.5%). The signals obtained by the sensors (sampling frequency = 1 KHz) are acquired by a data acquisition system and subsequently analyzed in a computer. Figure 7 shows the jet emanating from the nozzle during the experiments.

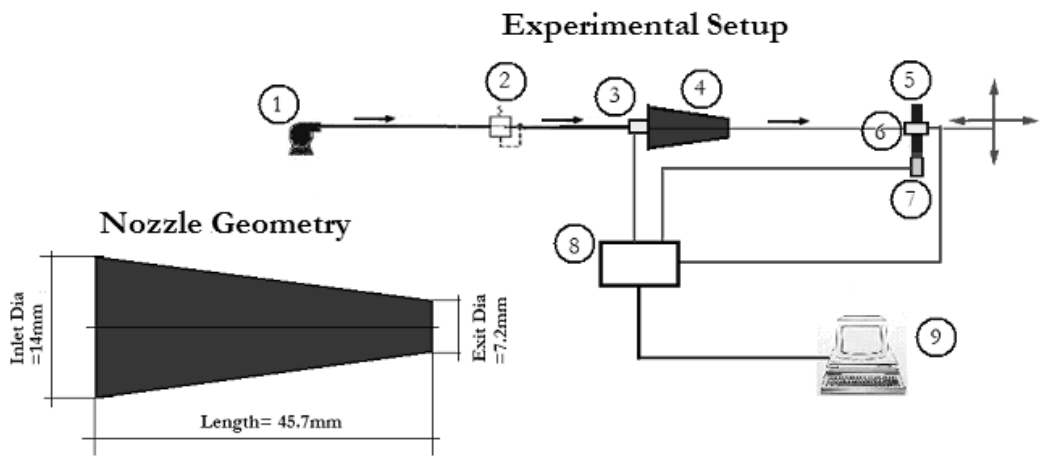

Figure 6: Schematic of the experimental set-up and the nozzle geometry. Labels : (1) Pump – 5 MPa, 0.00631 m$^3$/s  (2) Pressure Reducing Valve - up to 1.03 MPa (3) Pressure Transducer – 0 to 6.9 MPa, 0 to 5 V (4) Converging Nozzle (5) Target Plate (6) Pressure Transducer – 0 to 13.8 MPa, 0.468 to 10.397 V, 0.725 V/MPa (7) Linear Variable   Displacement Transducer – ±0.381m, 0-5 V, 0.0381 m/V (8) A/D Converter with 8 channels (9) Computer

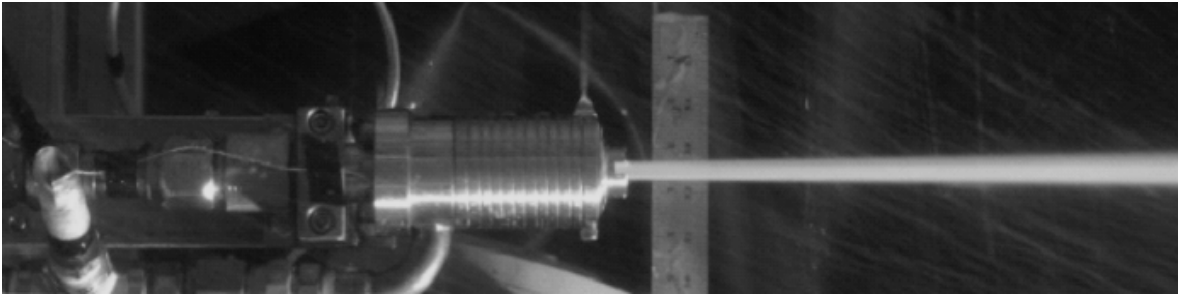

Figure 7: High speed water jet for cleaning operations.

Table 1 lists the different test cases considered in this experiment. For all the cases, the target plate is moved along the centreline with the aid of a robotic arm from 0.085 m to 0.310 m from the nozzle exit.

| Test Case | Nozzle Inlet Pressure ($P_N$) | Mass flow rate ($\dot{M}$) | Nozzle exit velocity ($V_o$) | Spread Coefficient ($C$) |
|---|---|---|---|---|
| | MPa | Kg/s | m/s | |
| 1 | 2.07 | 3.096 | 67.4 | 0.028 |
| 2 | 3.10 | 3.796 | 82.6 | 0.037 |
| 3 | 4.96 | 4.613 | 104.0 | 0.056 |

Table 1: Different experimental test cases

The starting point is kept at 0.085 m from the nozzle exit in order to ensure that the main stream jet flow does not get obstructed by the rebounding flow from the target plate. Also within 0.085 m, it is very difficult to take correct measurements of the Spread Coefficient, $C$. In fact, the rebounding flow can cause damage to the pump and accessories. The velocity of the robotic arm is 0.01 m/sec and thus the relative motion between the jet and the robotic arm is negligible. The target pressure and axial displacement data are recorded during this operation. In this way, the distribution of pressure along the centreline of the jet is obtained. In the next step, the target plate is kept at a fixed axial distance of 0.3098 m from the nozzle exit and is moved radially. The target pressure and radial displacement data were recorded during this operation.

The Spread Coefficients ($C$) of the jets for different test cases are obtained with the help of a scale (1 mm resolution) and photographs captured by a Nikon D300 camera. The former is defined as the ratio of difference between the radial width of the jet at the location of the scale and the nozzle exit radius ($R_N$) to the axial distance between the scale and the nozzle exit. Once $C$ is known, $k_1$ can be estimated using Equation 5.

The centreline pressure on the target plate, $P_T(x,0)$ (which is the stagnation pressure) is found to vary linearly with the axial distance. On normalizing the target pressure by total pressure at the nozzle inlet ($P_N$) and normalizing the axial distance by nozzle exit diameter ($D = 0.0072$ m), it is found that all the pressure curves collapse onto one curve. This is shown in Figure 8. This linear relation can be expressed as

$$\frac{P_T\left(\frac{x}{D},0\right)}{P_N} = m \times \left(\frac{x}{D}\right) + \frac{P_E}{P_N} \tag{14}$$

where $m$ is the slope, and is -0.0127 under the experimental conditions. $P_E$ is the pressure at the nozzle exit. In industrial cleaning nozzles, the losses inside the nozzle due to friction is negligible, hence the rightmost term in the right hand side is almost equal to 1. It is worth mentioning that this linear relationship holds for distances at least within ~50$D$.

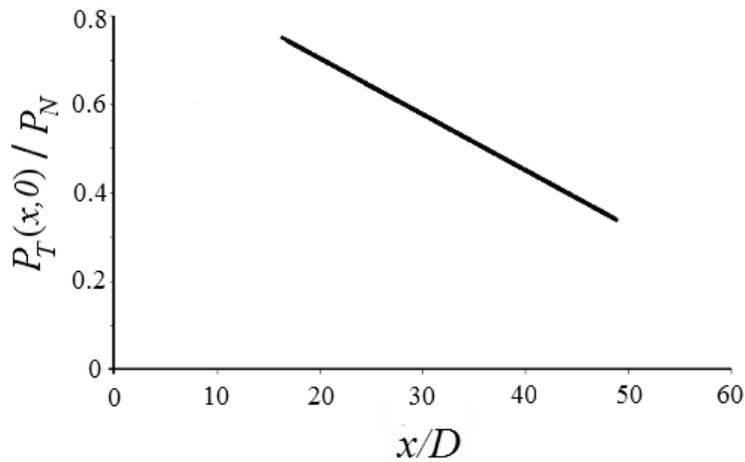

Figure 8: Normalized stagnation pressure on the target plate.

In the next experiment, the target plate is kept at a fixed stand-off distance of 0.3098 m from the nozzle exit and is moved radially. Figure 9 shows the radial distribution of target pressure for all the three cases, after normalizing the pressure ($P_T$) by the corresponding pressure on the jet axis ($P_T(x/D,0)$) and radial distance by nozzle exit diameter. The normalized distributions for these test cases are self similar and can be expressed by the following Gaussian distribution:

$$\frac{P_T\left(\frac{x}{D},\frac{r}{D}\right)}{P_T\left(\frac{x}{D},0\right)} = \exp\left(-2.8345\left(\frac{r}{D}\right)^2\right) \tag{15}$$

The radial position ($R_{Patm}$) corresponding to the value zero of the Gaussian curve fit is the position where the pressure on the target plate is equal to the atmospheric pressure. The value of $R_{Patm}$ is found to be 1.68$D$. The deviation with the experimental results of Leach et al. (1966) is clearly visible. They predicted $R_{Patm}$ to be 1.3$D$. The difference with our result is because their prediction was based on an analytical curve which is a third order polynomial satisfying the boundary conditions of the given problem. Our Gaussian curve fit has an R-squared value ~ 0.97. This is the reason why our numerical results in Figure 5 deviated from the experimental results of Leach et al. (1966). It is interesting to note that the value of $R_{Patm}$ is invariant with the axial position of the target plate. Examining Equation (15) we observe that the exponential part which represents the distribution of the pressure on the target plate is only dependent upon the radial coordinate. Thus, although the jet spreads in air and its radial width increases, the radial location where the target pressure turns atmospheric ($R_{Patm}$) remains fixed at the value of 1.68$D$ (See Figure 10).

Figure 10 shows the schematic of the pressure distribution on the target plate at two different locations, Location 1 being nearer to the nozzle. The pressure is maximum at the centreline and its distribution is Gaussian in the radial direction. As the jet progresses, it transfers momentum to the surroundings and thus continues to spread. This results in not only a reduced peak pressure (stagnation pressure) at the target plate, but also a decrease in overall pressure distribution.

Combining Equations 14 and 15, the overall pressure characteristics at the target plate can be represented as follows

$$P_T\left(\frac{x}{D}, \frac{r}{D}\right) = P_N\left[m \times \left(\frac{x}{D}\right) + \frac{P_E}{P_N}\right] \times \exp\left(-2.8345\left(\frac{r}{D}\right)^2\right) \quad (16)$$

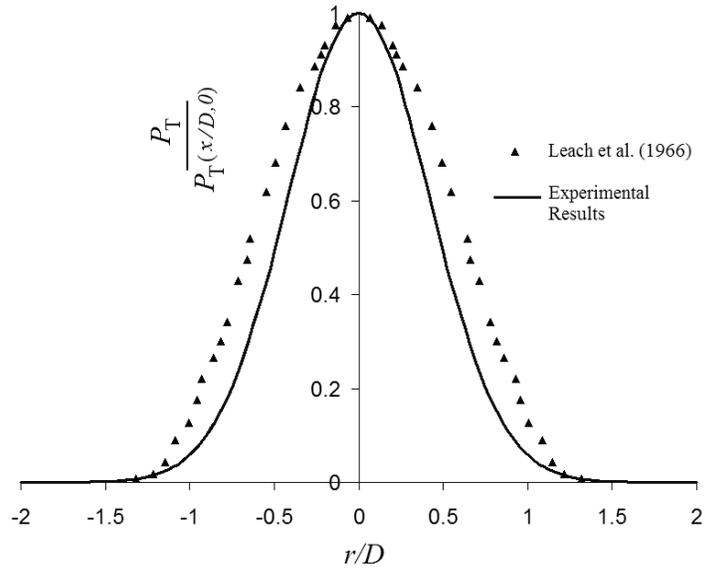

Figure 9: Experimental normalized target pressure along the radial direction and its comparison with Leach et al. (1966).

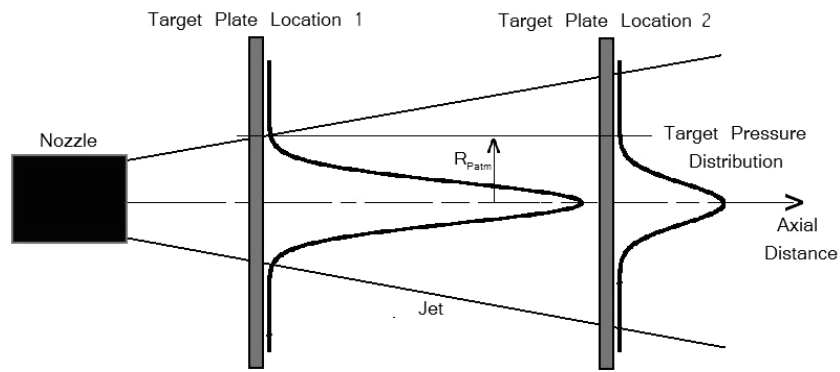

Figure 10: Schematic of the pressure distribution on the target plate.

In Springer (1976), investigation of the repetitive impacts of multiple liquid droplets on a solid surface was performed. They proposed that material removal occurs due to fatigue at a certain number of stress cycles when the equivalent dynamic stress is between the ultimate strength and the endurance limit of the material. Material removal ceases to occur if the equivalent dynamic stress is smaller than the material's endurance limit. This can be mathematically represented as follows:

$$\dot{M}\lambda\psi \geq S \tag{17}$$

Here $\lambda$, $\psi$ and $S$ are the stress coefficient (depending on the droplet size, coating thickness, and properties of the liquid, coating, and substrate material), speed of sound in liquid and endurance limit respectively. Based on this theory, Leu et al. (1998) proposed a critical stand-off distance, $x_c$ (shortest stand-off distance at which no cleaning is possible) and its relation with the optimal stand-off distance, $x_m$ (stand-off distance providing maximum cleaning). It occurs when the following condition is satisfied:

$$\dot{M}(x,0)\lambda\psi = S \tag{18}$$

From that they found the following relations:

$$x_c = 2.82\left(\frac{\lambda\psi k}{S}\right)^{0.5}\left(\frac{R_N}{C}\right)\left(\frac{P_N}{\rho_{w0}}\right)^{0.25} \tag{19}$$

$$x_m = 0.576 x_c \tag{20}$$

Here $k$ is the flow resistance coefficient and is ~0.96-0.99. Although their derivation methodology was right, we have noted a vital shortcoming of this approach. . They derived Equations (19) and (20) based on Equation (4). Evaluation of critical or optimal stand-off distance should be based on Equation (3) rather than Equation (4). This is because stand-off distance is evaluated along the centreline, hence Equation (3), which deals with the inner region of the jet is the appropriate one. Incorporating this change and proceeding much like Leu et al., the new formula for critical stand-off distance is found to be:

$$x_c = \left(\frac{R_N}{k_1^2}\right)\left(2.37\sqrt{\frac{\rho_{w0}V_{w0}\lambda\psi}{S}} - 1\right)^2 \tag{21}$$

Also, we are able to obtain a relation between the critical and optimal stand-off distances which is as follows:

$$0.576\sqrt{x_c} - \sqrt{x_m} = 0.424\left(\frac{\sqrt{R_N}}{k_1}\right) \quad (22)$$

If the Right hand side of Equation (22) is small, it can be approximated to be zero. In that case,

$$x_m \approx 0.33 x_c \quad (23)$$

As pointed out in Leu et al. (1998), it is extremely difficult to evaluate $\lambda$, hence these theoretical relations are not practically helpful in evaluating $x_c$ or $x_m$. Also, we found (see Figure 8) that $x_m$ (or $x_m/D$) cannot be measured experimentally because keeping the target plate too close to the nozzle exit produces flow reversal. Fortunately, our numerical simulations have given accurate predictions of jet characteristics; hence it can be applied to evaluate $x_m$.

## 6. Numerical Simulation

Corresponding to the experimental results obtained in Section 5, numerical simulations are performed. Since we didn't perform any velocity measurements, the velocity profile at the nozzle exit is not known to us. We have measured the mass flow rate and from that we calculated the average velocity at the nozzle exit. Since the nozzle inlet pressure is known and the nozzle outlet pressure is atmospheric, we can use these parameters as well as the mass flow rate to numerically simulate the flow features inside the nozzle. For our nozzle geometry, we have found that the flow inside the nozzle becomes fully developed turbulent flow. Under the assumptions that the turbulence intensity at the nozzle inlet is 10% and the nozzle is frictionless, we obtained the nozzle exit velocity to be of top-hat profile. Figure 11 shows the velocity and turbulence intensity distribution in the nozzle for the Test Case 2 (the corresponding simulation is named Sim 2). This nozzle exit velocity and turbulence profile is used as the boundary condition for the jet flow problem.

The domain is made 309.8 mm × 152.4 mm. The boundary conditions and other relevant parameters are already known from the experimental conditions. The methodology of creating the domain, setting up the CFD solver and incorporation of the semi-empirical mass flux model in FLUENT is the same as that discussed in Section 3.

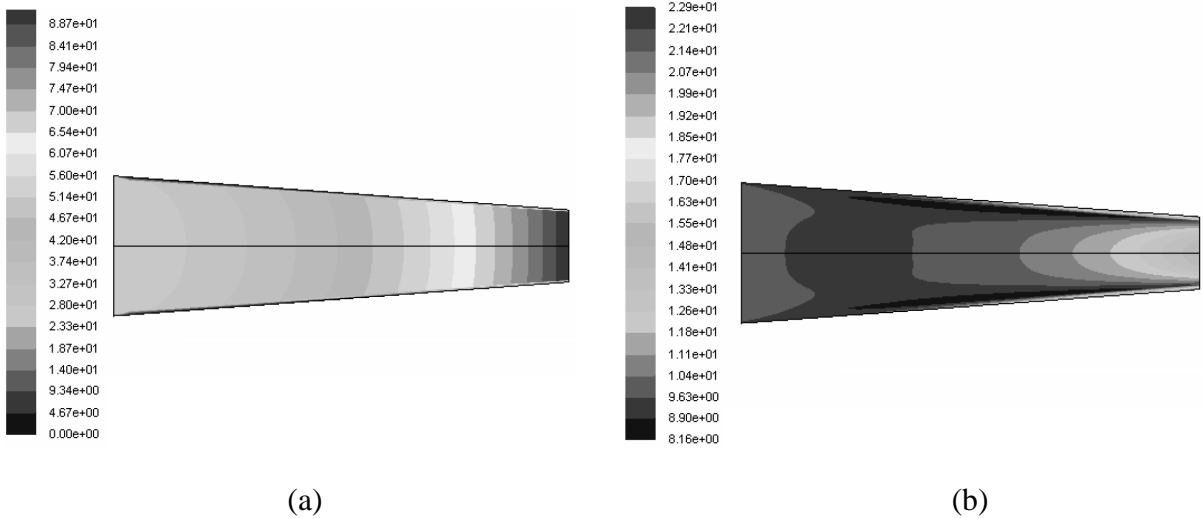

(a) (b)

Figure 11: (a) Velocity distribution and (b) Turbulence Intensity distribution inside the nozzle corresponding to Test Case 2 (i.e. Sim 2).

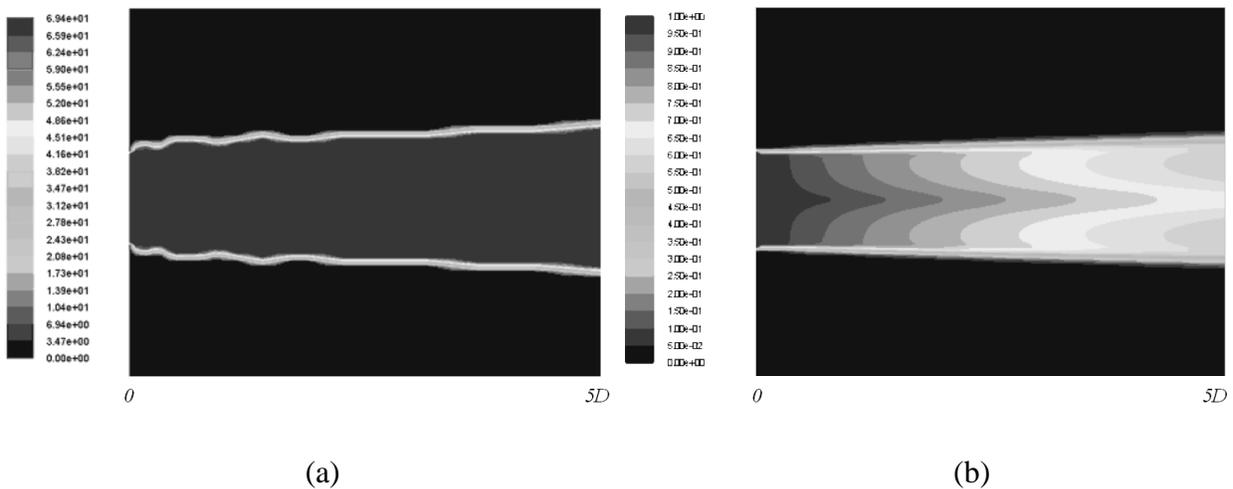

(a) (b)

Figure 12: (a) Velocity and (b) Volume fraction of water phase of the jet corresponding to Test Case 1 (i.e. Sim 1).

Figure 12 shows the velocity and volume fraction contours for Sim 1. The contour plot shows the behavior of the jet after it emanates from the nozzle (the maximum axial distance shown in the figure is 5$D$). Several qualitative inferences can be drawn from Figure 12, viz. (i) the velocity profile is top-hat, (ii) there are undulations on the velocity profile which clearly show the growth of instability in the shear layer as one would expect, (iii) the jet expands almost linearly, (iv) a strong shear layer is concentrated at the jet boundary and (v) considerable amount of air is entrained within the jet and the water phase volume fraction is maximum at the centerline and decreases rapidly with radius. Although the results here are for Sim 1, exactly similar behavior is observed in the other simulation cases.

Figure 13 (a) and (b) respectively shows the water phase velocity and volume fraction distributions in the radial direction for three different axial distances. The case considered here is Sim 1. As has already been discussed, the velocity distribution remains top hat even at 30*D* mainly due to the sharp density interface (~1:815) between air and water. The volume fraction distribution shows a kind of Gaussian profile with a bulge. Since the mist region is not included in the numerical modeling, the volume fraction of water actually lost as mist numerically accumulates near the jet-air interface and produces the erroneous bulging effect. The bulging effect flattens out with increased axial distance. It is seen that the volume fraction decreases rapidly with increased axial distance while the velocity remains fairly constant. Also the volume fraction profile is thinner than the velocity profile. All these observations are in congruence with the experimental findings of Rajaratnam et al. (1998).

Figure 14 shows the comparison between the experimental and numerical results of the pressure distribution on the target plate for the three test cases. The stand-off distance is kept constant at 0.3098m (x/*D* = 43.03). The difference between the experimental and numerical results is primarily due to the uncertainty in measuring *C*, which is ~7-8%.

In order to find the optimal stand-off distance, a set of simulations are performed by moving the target plate boundary. It is found that the stagnation pressure ($P_T(x,0)$) is maximum at *x*=5*D*, and there is no backward flow (in case of backward flow, the solution diverges). This is intuitively right, since ~5*D* marks the end of the potential core. Thus, on the basis on Equation (22) and our experimental parameters, the optimal and critical stand-off distances can be summarized as follows:

$$x_m = 5D , x_c = 26D \tag{24}$$

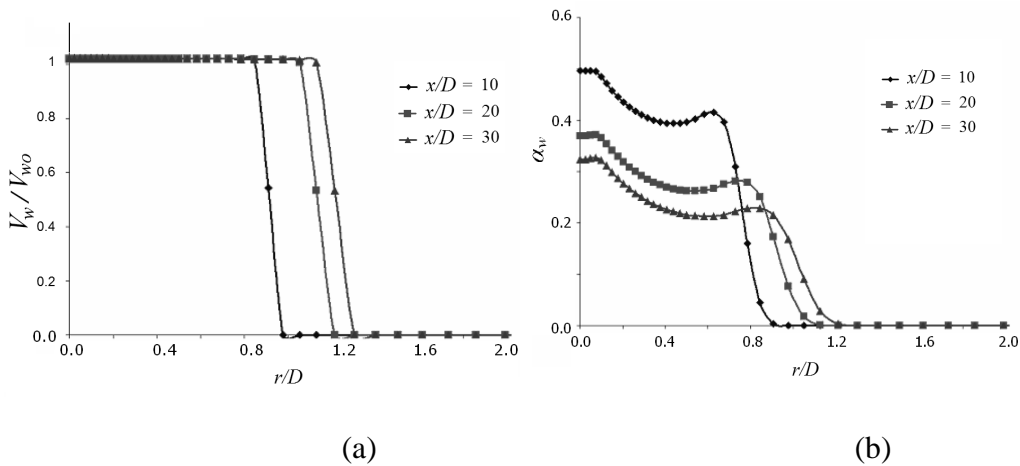

(a)          (b)

Figure 13: (a) Water phase velocity and (b) Volume Fractions at x/*D*=10, 20 and 30 corresponding to Sim 1.

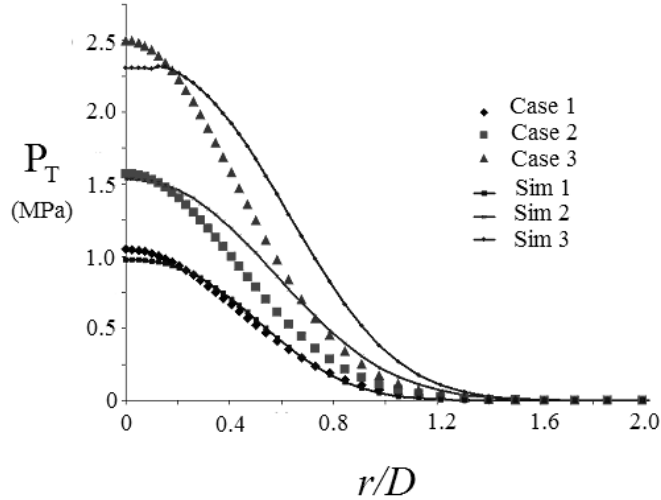

Figure 14: Comparison of experimental (denoted by 'Case') and corresponding simulation (denoted by 'Sim') results for target pressure

## 7. Conclusion

Both experimental study and numerical simulations are performed to understand the behavior of cleaning jets. These jets diffuse in the surrounding atmosphere by the process of mass and momentum transfer. Air is entrained into the jet stream and the whole process contributes to the spreading of the jet and subsequent decay of pressure. The magnitude and distribution of pressure on the cleaning surface (target plate) has been obtained using both experimental and numerical methods. A semi-empirical method has been implemented in the numerical simulation to capture the air entrainment process properly. The numerical results are validated against the experimental results available in literature as well as our own experiments. Also, a theoretical model to calculate the critical and optimal stand-off distances has been proposed. Based on our numerical results, we have concluded that the optimal stand-off distance is ~$5D$ from the nozzle exit, and the jet looses its cleaning ability at ~$26D$ (critical stand-off distance). Keeping the target plate too close to the nozzle causes the jet to rebound from the cleaning surface and obstruct the nozzle flow, thereby decreasing cleaning efficiency. On the other hand, if the surface is kept too far from the nozzle, the jet will transfer momentum to the surroundings, thereby producing huge pressure loss and thus inefficient cleaning. The latter is a direct consequence of the excessive water phase volume fraction decay (Figure 3).

From our experimental findings, we have found that the cleaning radius ($R_{Patm}$) of the jet is $1.68D$. Thus, debris located at a radial distance greater than $1.68D$ is definitely irremovable by the cleaning jets.

Our proposed simulation methodology can be helpful for predicting the flow behavior of jets used in industrial cleaning applications since these applications focus on the near-field region of the jet. The simulation correctly predicts the magnitude and distribution of pressure on the cleaning surface. The knowledge of nozzle inlet conditions and the Spread Coefficient are required as parameters for the simulation, and these parameters can be obtained without much difficulty. The experimental and numerical results on pressure distribution on the target plate will help the understanding of the erosion process and subsequent material deformation.

Finally, it is worth mentioning that the experimental and numerical methodologies can be extended to water jet cutting operations.

## Acknowledgement

We would like to thank Valiant Machine and Tool Inc., Windsor, Ontario, Canada for providing us experimental support.

## References

Adler, W. F., 1979. The Mechanisms of Liquid Impact. Treatise on Materials Science and Technology, Academic Press.

FLUENT 6.3.26 User Manual

Guha, A., Barron, R. M., and Balachandar, R., 2010. Numerical Simulation of High Speed Turbulent Water Jets in Air. Journal of Hydraulic Research, 48(1), 119-124.

Hashish, M., and duPlessis, M. P., 1978. Theoretical and Experimental Investigation of Continuous Jet Penetration of Solid. ASME Journal of Engineering for Industry, 100, 88–94.

Hashish, M., and duPlessis, M. P., 1979. Prediction Equations Relating High Velocity Jet Cutting Performance to Standoff Distance and Multipasses. ASME Journal of Engineering for Industry, 101, 311–318.

Leach, S. J., Walker, G. L., Smith, A. V., Farmer, I. W., Taylor, G., 1966. Some Aspects of Rock Cutting by High Speed Water Jets. Philosophical Transactions of the Royal Society of London, 260 (1110), 295-310.

Leu, M.C., Meng, P., Geskin, E.S., Tismeneskiy, L., 1998. Mathematical Modeling and Experimental Verification of Stationary Water Jet Cleaning Process. Journal of Manufacturing Science and Engineering, 120(3), 571-579.

Lin, S.P., Reitz, R.D., 1998. Drop and Spray Formation from a Liquid Jet, Annual Review of Fluid Mechanics, 30, 85–105.

Liu, H.,Wang, J.,Kelson, N.,Brown, R.J., 2004. A study of abrasive waterjet characteristics by CFD simulation. Journal of Materials Processing Technology, 153-154, 488-493.

Rajaratnam, N., Steffler, P.M., Rizvi, S.A.H., Smy, P.R., 1994. Experimental Study of Very High Velocity Circular Water Jets in Air. Journal of Hydraulic Research, 32(3), 461-470.

Rajaratnam, N., Albers,C., 1998. Water Distribution in Very High Velocity Water Jets in Air. Journal of Hydraulic Engineering, 124(6), 647-650.

Springer, G. S., 1976. Erosion by Liquid Impact. Scripta Publishing Co., Washington, DC.

Watson, J. D., 1993. Thermal Spray Removal with Ultrahigh-Velocity Waterjets, Proceedings of 7th American Waterjet Conference, Seattle, WA, pp. 583–598.

Yoon, S.S., Hewson, J.C., DesJardin, P.E., Glaze, D.J., Black, A.R., Skaggs, R.R., 2004. Numerical Modeling and Experimental Measurements of a High Speed Solid-Cone Water Spray for Use in Fire Suppression Applications, International Journal of Multiphase Flow, 30, 1369–1388.

## Appendix

The User Defined Program (UDF) used to validate the experimental results of Rajaratnam et al. (1994, 1998) is as follows:

```
#include "udf.h"
#include "threads.h"
#include "metric.h"
#include "mem.h"
#include "sg_mphase.h"

DEFINE_MASS_TRANSFER(watertoair,cell,t,from_index,from_species_index,to_index,to_species_index)
{
/* variable declarations*/
real m_lg1;
real x[ND_ND];
double Ro,Ri,rad;
real rN=0.001; /*radius of nozzle=1mm*/
real v0=155;/*INPUT velocity*/
```

```
real termRo,termRi;
double dMdrRo,dMdrRi,dMdRo1,dMdRo2,dMdRo,dMdRi1,dMdRi2,dMdRi,dRodx,dRidx;
real const1=v0*rN*rN*5.62*998.2;/*alpha × rho × vel ×  rN ×  rN  ×  5.62 */
m_lg1=0; /*mass transfer term as reconginzed by FLUENT */
C_CENTROID(x,cell,t);
Ro=0.05*x[0]+ rN;/*Ro is the width in the outer region.. Eq 4*/
Ri=0.65*sqrt(rN*x[0])+rN;/*Ri is the width in the inner region.. Eq 3*/
rad=x[1]; /*radial coordinate*/

/* defining the mass flow rate derivatives in the two regions */
dMdrRo=4.5*const1*pow((1-pow(rad/Ro,1.5)),2)*pow(rad/Ro,0.5)*(1/(Ro*Ro*Ro));
dMdrRi=4.5*const1*pow((1-pow(rad/Ri,1.5)),2)*pow(rad/Ri,0.5)*(1/(Ri*Ri*Ri));

dMdRo1=4.5*const1*rad*pow((1-pow(rad/Ro,1.5)),2)*pow(rad/Ro,0.5)*(1/(Ro*Ro*Ro*Ro));
dMdRo2=-2*const1*pow((1-pow(rad/Ro,1.5)),3)/(Ro*Ro*Ro);
dMdRo=dMdRo1+dMdRo2;

dMdRi1=4.5*const1*rad*pow((1-pow(rad/Ri,1.5)),2)*pow(rad/Ri,0.5)*(1/(Ri*Ri*Ri*Ri));
dMdRi2=-2*const1*pow((1-pow(rad/Ri,1.5)),3)/(Ri*Ri*Ri);
dMdRi=dMdRi1+dMdRi2;

dRodx=0.05;/* derivative of Eq 4 wrt x*/
dRidx=0.325*rN/sqrt(rN*x[0]);/* derivative of Eq 3 wrt x*/

termRo=dMdrRo+dMdRo*dRodx;
termRi=dMdrRi+dMdRi*dRidx;

if(rad<=Ri) /*if within the inner region of the jet.. Eq 2 */
{m_lg1=termRi/4000;} /*..the number 1/4000 comes from the cell dim */
if((rad<=Ro)&&(rad>Ri)) /*if within the outer region of the jet.. Eq 2*/
{m_lg1=termRo/4000;}
if(rad>Ro) /*neglect mass transfer in mist region*/
{m_lg1=0;}
```

```
return m_lg1;
}
```

## List of Tables

## List of Figures

Figure 13: (a) Water phase velocity and (b) Volume Fractions at x/*D*=10, 20 and 30 corresponding to Sim 1.

Figure 14: Comparison of experimental (denoted by 'Case') and corresponding simulation (denoted by 'Sim') results for target pressure